\documentclass{article}
\usepackage{spconf,amsmath,graphicx}
\usepackage{bm}
\usepackage{multicol}
\usepackage{multirow}
\usepackage{amssymb}
\usepackage{subfigure} 
\usepackage{cite}
\usepackage{xcolor}
\ninept
\newcommand{\edited}[1]{\textcolor{black}{#1}}

\title{SSLIDE: Sound Source Localization for Indoors based on Deep Learning}
\name{Yifan Wu, Roshan Ayyalasomayajula, Michael J. Bianco, Dinesh Bharadia, and Peter Gerstoft}
\address{University of California, San Diego, La Jolla, CA, USA}
\begin{document}
%\ninept
%
\maketitle
\begin{abstract}
This paper presents \textit{SSLIDE}, Sound Source Localization for Indoors using DEep learning, which applies deep neural networks (DNNs) with encoder-decoder structure to localize sound sources with random positions in a continuous space. The spatial features of sound signals received by each microphone are extracted and represented as likelihood surfaces for the sound source locations in each point. Our DNN consists of an encoder network followed by two decoders. The encoder obtains a compressed representation of the input likelihoods. One  decoder resolves the multipath  caused by reverberation, and the other decoder  estimates the source location.  Experiments based on both the simulated and experimental data show that our method can not only outperform multiple signal classification (MUSIC), steered response power with phase transform (SRP-PHAT), sparse Bayesian learning (SBL), and a competing convolutional neural network (CNN) approach in the reverberant environment but also achieve a good generalization performance. 

\end{abstract}
\section{Introduction}\label{sec:intro}
Sound source localization (SSL) has widespread applications in human–robot interaction \cite{nakadai2000active}, ocean acoustics \cite{gemba2019robust}, teleconferencing \cite{zhao2012real}, and automatic speech recognition \cite{xiao2014ntu}. For example, in a hospital, attending robots can locate and attend to patients based on their voices \cite{argentieri2015survey}. However, SSL in reverberant environments is challenging due to multipath artifacts in received signals. This effect degrades SSL performance. Thus, it is important to develop SSL methods that are robust to reverberation\cite{bianco2019machine}.

% Conventional SSL approaches rely on estimation theory \edited{[REFERENCE]} and second-order statistics \edited{[REFERENCE]}. A well-known subspace based technique, multiple signal classification (MUSIC) \cite{schmidt1986multiple} 
% can be used to estimate the direction-of-arrival (DOA) of the sound source.  MUSIC is known to fail for interfering sources, as in a reverberant room\edited{[REFERENCE]}.  Another classical SSL method, steered  response  power  with  phase  transform  (SRP-PHAT) \cite{dmochowski2007generalized, velasco2016proposal, yook2015fast}, is also adopted in SSL. Although SRP-PHAT was developed to be robust to reverberation, it has been shown to not be robust to signal non-stationarity (e.g. speech).

\edited{While traditional SSL algorithms \cite{dibiase2001robust, schmidt1986multiple,dmochowski2007generalized,velasco2016proposal, yook2015fast, gustafsson2003source} rely on estimation theory or statistics, they fail in dynamic and reverberant environments. A well-known subspace based technique, multiple signal classification (MUSIC) \cite{schmidt1986multiple} is known to suffer from correlated sources which are prevalent in reverberant environments. Another classical SSL method, steered  response  power  with  phase  transform  (SRP-PHAT) \cite{dmochowski2007generalized, velasco2016proposal, yook2015fast} has been shown to not be robust to non-stationary signal like speech.} Recently, SSL approaches based on deep neural networks (DNNs) have been proposed \cite{xiao2015learning, chakrabarty2019multi, takeda2016sound, adavanne2018sound, wang2018robust, takeda2017unsupervised, laufer2016semi,bianco2020semi}. Most of the approaches are based on supervised learning. In \cite{xiao2015learning}, a multilayer perceptron DNN is proposed for DOA estimation. In \cite{chakrabarty2019multi}, a SSL framework based on convolutional neural network (CNN) is proposed. A learning based SSL approach using discriminative training is presented in \cite{takeda2016sound}. The authors in \cite{adavanne2018sound} propose a convolutional recurrent DNN for SSL and sound event detection. In \cite{wang2018robust}, a robust SSL guided by deep learning based time-frequency masking framework was presented. There are also some works using unsupervised learning \cite{takeda2017unsupervised} and semi-supervised learning methods based on manifold learning \cite{laufer2016semi}, and deep generative modeling \cite{bianco2020semi}. But all of these methods can only work well when the sensor-source distance is small, which limits their implementation in real-world settings. 

In this work, we present \textit{SSLIDE}, a SSL method based on DNN with encoder-decoder structure. Our method can resolve randomly located sources in the room and can achieve a good generalization performance. Inspired by \cite{ayyalasomayajula2020deep}, the major novelty of the architecture lies in the two parallel decoders \edited{that help in solving two distinct and independent problems}. One decoder is designed to resolve the multipath artifacts, and the other to predict the locations of the sound sources. By training these decoders in parallel, the  DNN learns to jointly predict the locations of sound sources and remove the multipath artifacts on range offsets. We compare our approach with other baseline SSL methods, including multiple signal classification (MUSIC)\cite{schmidt1986multiple}, steered response power with phase transform (SRP-PHAT) \cite{dmochowski2007generalized, velasco2016proposal, yook2015fast}, Sparse Bayesian Learning (SBL) \cite{xenaki2018sound}, and CNN \cite{chakrabarty2019multi}. Based on the experiment results, we find SSLIDE outperforms  the baseline methods and generalizes well across space, perturbations of reverberation time, microphone spacing, and input speech.

\section{Proposed Method}\label{sec:format}
%\subsection{Data model}
\edited{To understand how the proposed DNN solves for the reverberation problem and helps in efficient SSL, let us first look into the fundamentals of sound transmissions in a given environment.} Consider the acoustics signals in the time domain
\begin{equation}
    y_{i} = s * h_{i} + n_{i} 
\end{equation}
where $y_{i} \in \mathbb{R}^{L}$ is the signal received by $i$th microphone ($i \in \{1, \ldots, M\}$, $M$ is the number of microphones), $s$ the source signal, and $n_{i}$ the noise for the $i$th microphone. $h_{i}$ is  the room impulse responses (RIRs), which characterizes the reverberation of the room. Denote $\textbf{y} = [y_{1},  \ldots, y_{M}]^{T} \in \mathbb{R}^{M \times L}$ as the collections of the received signal of all sensors with audio length $L$. 

For $N$ arrays with $K$ microphones in each array (i.e. $M = NK$), $\textbf{y}$ can be reshaped as a tensor with dimension $K \times N \times L$. So, for a given input received signal $y$  with $S$ snapshots and $T$ datapoints (number of independent measurements), there are $C = TS$ frames for $y$. 

\subsection{Features extraction} 
\label{features}
\edited{One of the key components in designing a DNN model is  to understand the input data and represent it appropriately for the network to be able to learn the required application using the input data. While there have been existing works like MUSIC and SRP-PHAT that enable accurate SSL for environments with low reverberation, their SSL performance degrades significantly in dynamic and reverberant environments.} 

We first use standard beamforming to obtain source location likelihoods in a 2D space. We obtain $\textbf{Y} \in \mathbb{C}^{S \times F \times K \times N}$, the STFT output of $\textbf{y} $ with $S$ snapshots and $F$ frequency bins\edited{, where $L = 2FS$. While $2F$ is the total number of frequency bins in the STFT, we only consider the positive half of the frequency bins}. For $T$ datapoints, then there are $C = TS$ frames for $\textbf{Y}$.

Assume a uniform linear array (ULA) and a broadband signal. Inspired by \cite{ayyalasomayajula2018bloc}, we can define a 2-D function \cite{ayyalasomayajula2018bloc} which can indicate the likelihood of the signal coming from the angle $\theta$ and distance $d$ for array $n \in \{1, ..., N\}$ and frame $s \in  \{1, ..., S \}$
\begin{equation}
    P_{ns}(\theta, d) = |\sum_{i=1}^{K} \sum_{l=1}^{F} Y_{il} e^{j\frac{2 \pi i u \text{sin} \theta f_{0}}{c}} e^{j \frac{2 \pi l f_{l} d}{c}}|
\end{equation}
where $j = \sqrt{-1}$ and $u$, $f_{0}$, $f_{l}$, $c$ stand for the spacing between microphones, median frequency, the frequency corresponding for the $l$th frequency bin, the speed of sound, respectively. $P_{ns}(\theta, d)$ is beam power. $Y_{il}$ represents the STFT output for the $i$th microphone and $l$th frequency bin. When the sound source is from  angle $\theta$ and distance $d$, then $P_{ns}(\theta, d)$ have a high value. If we have $U$ and $V$ grid points for $\theta$ and $d$, then we will obtain a likelihood surface with dimension $U \times V$ which can indicate the likelihood of the signal in the given $\theta$ and $d$. Fig. \ref{eg} (a) is one of the examples.

For reverberation and noise free data, the localization is simply  identifying the $\theta$ and $d$ that correspond to the maximum likelihood \cite{ayyalasomayajula2020deep}. Due to the reverberation, much of the sound received by the microphones is a result of multipath, which is a complicated function of the different microphone locations relative to the source. Therefore, peaks in the likelihood surface may no longer indicate the correct result \edited{in terms of their predicted distance $d$ as depicted in Fig. \ref{eg} (a)}.

\subsection{Range compensation}
\label{range}
To help overcome challenges of source localization in reverberant environments, we design a second decoder to explicitly correct for variation in multipath artifacts due to differences in microphone location. Details on the decoder and the loss function are further described in Section 2.4.

\begin{figure}[t]
\centering
\includegraphics[width=8.5cm]{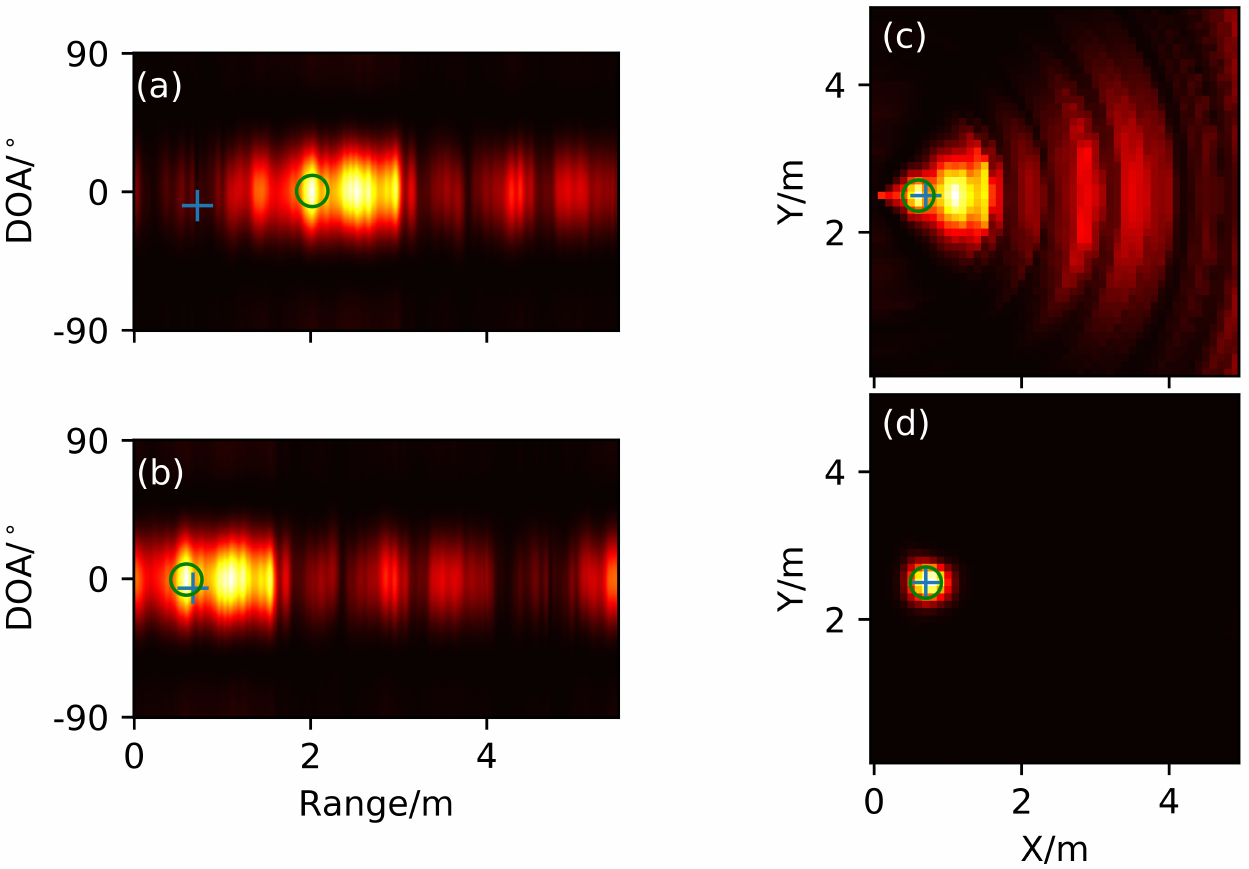}
\caption{Likelihood surfaces (a) before and (b) after range compensation, and (c) after range compensation and coordinate transformation,  (d) Output likelihood surface from localization decoder.  The correct source position (+), and maximum value (for (a) - (c)) / predicted location (for (d)) ($\circ$) are indicated. All plots are based on the same testing data.}
\label{eg}
\end{figure}

To enable this decoder to learn to alleviate range offsets cause by multipath artefacts, we will artificially generate likelihood surfaces with range compensation as labels. To do so, we first identify the direct path as the path with the least range measurement, $\hat{d}$ in the incorrect range image as shown in Fig. \ref{eg} (a). We then use the actual range measurement expected range measurement, $d$, from the given ground truth location for that specific measurement. We then compensate this offset in the given RIR measurement to get the expected likelihood profile as seen in Fig. \ref{eg} (b). More formally, for the STFT output in the $s$th frame and $k$th microphone of the $n$th array $Y^{ns}_{k} \in \mathbb{C}^{F \times 1}$, the range is compensated by
\begin{equation}
    \bar{Y}^{ns}_{k} = Y^{ns}_{k} \circ  e^{j2 \pi \vartheta \frac{\hat{d}_{n} - d_{n}}{c}} 
\end{equation}
where $\vartheta = [f_{1}, ..., f_{F}]^{T} \in \mathbb{R}^{F \times 1}$ is a collection of all frequencies. Scalar $d_{n}$ and $\hat{d}_{n}$ are the estimated ranges for the direct path and true ranges of the  $n$th array, and $\circ$ represents the Hadamard product. Fig. ~\ref{eg}(b) shows the likelihood surface after range compensation. Our results show that the range compensation will make DNN easier to identify the correct location of the sound source.

We have generated two categories of likelihood surfaces with dimension $U \times V$. \edited{While we can perform single point identification based object detection tasks on these images, each of these images are with respect to their own microphones and lack the context of the global coordinates. To overcome that problem, we convert these range-angle images into 2D Cartesian images which show the coordinate with respect to one of the arrays.} We perform a coordinate transform on these images to convert to a 2-D Cartesian plane with dimension $Y \times X$ as shown in Fig ~\ref{eg}(c) to encode the locations of these multiple arrays.

% Note that since we have already accessed to the ground-truth position in the range compensation process, this step can only be done during the training phase. 

\subsection{Target processing}
\label{target}
% \edited{ One naive way to generate the targets is to just mark the ground-truth $(x, y)$ coordinates as 1 and the rest of coordinates as 0, which might make the problem to regression which would make the learning more complicated and non-scalable to environments of different sizes. To avoid that we make ensure that t}he target of the network will also be the likelihood surfaces with dimension $Y \times X$ \edited{ and use only CNNs throughout the network architecture to avoid dense layers and thus making the network lightweight. }
Now that we have defined the images for us to perform the single point identification, we need to define the targets for the network to learn the SSL task. One naive way to generate the target images is to only mark the target position as one and the rest positions as zeros. Unfortunately, this method will make the loss extremely small, which will bring about the gradient underflows. The network cannot learn how to predict the locations due to the almost vanishing gradients. 

Thus, we use a negative exponential label to represent the target position. The target of the network will also be a likelihood surface with dimension $Y \times X$. The distance between a random position $(x', y')$ in the likelihood surface and ground-truth position $(x, y)$ is 
\begin{equation}
    d(x', y') = \sqrt{(x' - x)^{2} + (y' - y)^{2}};
\end{equation}
Then its value in the likelihood surface $l(x', y')$ will be marked as
\begin{equation}
\label{sigma}
    l(x', y') = e^{-d(x', y')^{2}/ \sigma^{2}}
\end{equation}
where $\sigma$ is a hyperparameter controlling the rate of decay. For $d(x', y') = 0$, then $l(x', y')=1$  its maximum value. Far from the target position, the value will decay significantly. For most of the points in the heatmap, the values is close to $0$. \edited{These output representations is helpful for a smoother gradient flow.}

\subsection{SSLIDE architecture}
\begin{figure}[t]
\centering
\includegraphics[width=8.5cm]{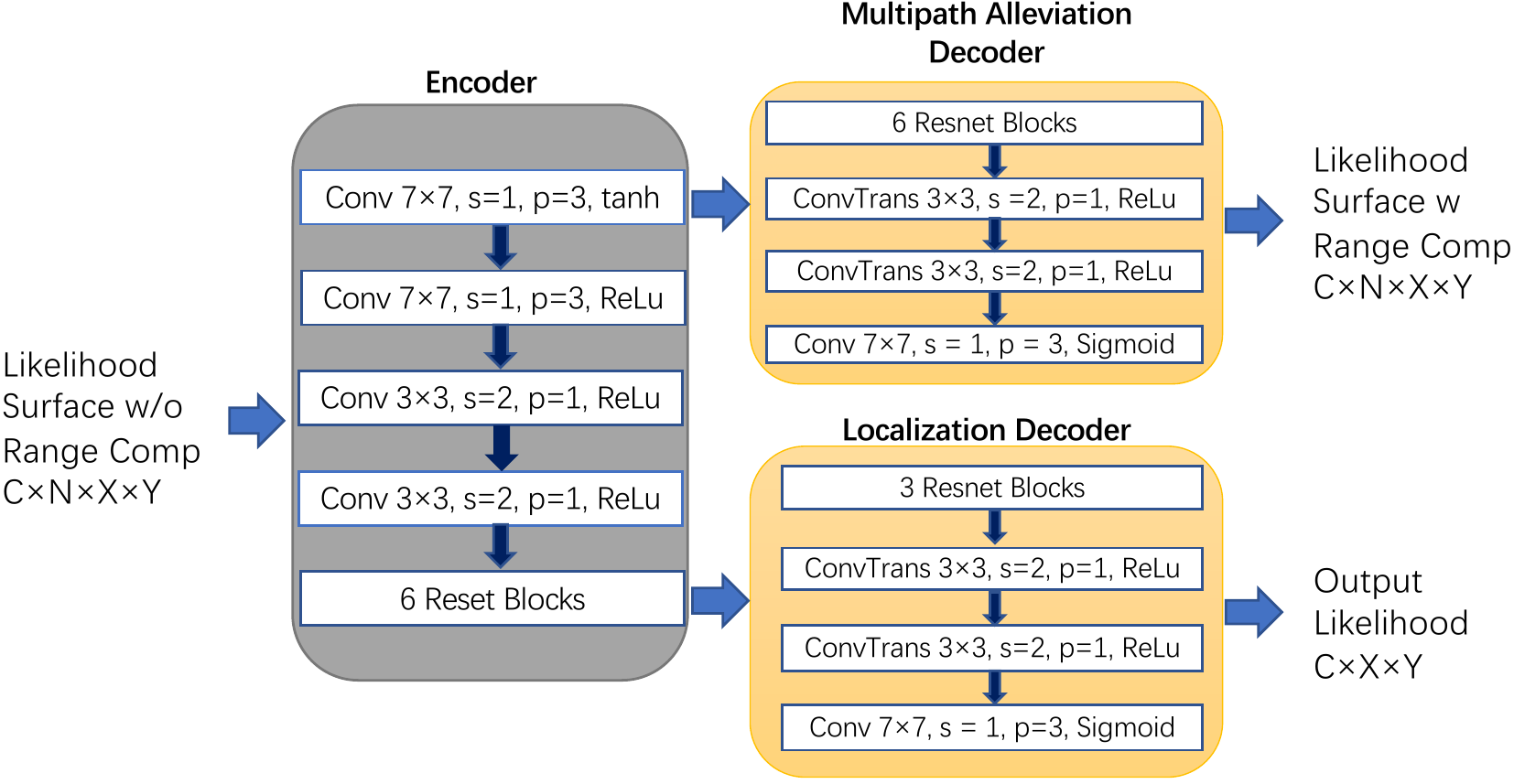} %{DNN_new.png}
\caption{SSLIDE architecture. $C$, $N$, $X$ and $Y$ are followed by the definitions in Sec 2.1--2.4. s and p stand for stride and padding, respectively}
\label{dnn}
\end{figure}

Now that we have the inputs and targets for performing single point identification, we utilize the network architecture as shown in Fig. \ref{dnn} and based on encoder-decoder architecture with one encoder and two parallel decoders inspired from \cite{ayyalasomayajula2020deep}. The input to the encoder is the likelihood surface without range compensation indicated as (2). The encoder compresses this representation and then feeds to both of the decoders. The two decoders will focus on two different tasks simultaneously. The multipath alleviation decoder will use the likelihood surface with range compensation mentioned in Sec. \ref{range} as targets to train the network to generate the likelihood surface without range offsets. With the help of this decoder, the neural network will learn the multipath profile and how to alleviate such an artifact with respect to the range estimation, which will facilitate the localization decoder to identify the source locations. The localization decoder aims to predict the location of the sound source by using the target likelihood surface mentioned in \ref{target} as labels. The output for the localization decoder is also a likelihood surface with dimension $Y \times X$. The location with the highest value in this output image will be marked as the predicted location. Note that since we have used the ground-truth position to generate the target images with range compensation, the multipath alleviation decoder will only appear during the training phase, and it will be turned off during the testing phase. 

% \newenvironment{figurehere}
% {\def\@captype{figure}}
% {}\begin{figurehere}
% \includegraphics[width=140mm]{DNN_2.png}

% \end{figurehere}

The loss function for the multipath alleviation decoder is $l_{2}$-loss 
\begin{equation}
    L_{multipath} = \frac{1}{N} \sum^{N}_{i = 1} \|I_{out}^{i} - I_{target}^{i}\|_{2}
\end{equation}
where $I_{out}^{i}$ and $I_{target}^{i}$ are the decoder outputs and the targets (likelihood surfaces with range compensation) of the $i$th array, separately. All of the outputs and targets are likelihood surfaces with the same dimension. $N$ is the number of arrays. \edited{The advantage of averaging across multiple receiver arrays is that we can enforce consistency of peaks across all the target images, and the network will learn the consistency across these multiple receiver arrays.}

For the localization decoder, we use $l_{2}$-norm loss with $l_{1}$ regularization to enforce the sparsity as \edited{there only exists one global maxima} in the output likelihood surface. The loss function of that decoder can be expressed as
\begin{equation}
\label{reg}
    L_{localization} = \|T_{out} - T_{target}\|_{2} + \lambda \|T_{out}\|_{1}
\end{equation}
where $T_{out}$ and $T_{target}$ are the decoder outputs and targets (target images with negative exponential labels), respectively. $\lambda$ is the regularization term. The loss functions from these two decoders are summed  and back-propagated to the input. Fig. \ref{eg} (d) shows the output likelihood surfaces from the localization decoder based on the same data as Fig. \ref{eg} (a).

\section{Experiments}
\label{sec:pagestyle}
The localization performance of SSLIDE and other baseline methods is evaluated with different levels of reverberation by using both simulated and real RIRs. The real RIRs are from Multi-Channel Impulse Responses Database (MIRD) \cite{hadad2014multichannel}. For each case, the networks for the learning-based methods are trained and evaluated on both datasets. 

\subsection{Datasets}
\subsubsection{Simulated Data}
%\textbf{Simulated Data}
Simulated RIRs are synthesized by the RIR generator \cite{rir2016}, which models the reverberation using the image method \cite{allen1979image}. The room   is $8 \times 5 \times 4 $ m with reverberation times ($RT_{60}$) from $0.2$--$0.8$ s and speed of sound $c = 340$ m/s. There are $N$ = 3 identical ULA with  array   centers  $(0, 2.5, 2)$, $(4, 0, 2)$, and $(8, 2.5, 2)$ m and $K$ = 4 sensors in each array with identical space $2.6$ cm.  To train the generalization across space, the sources have random $(x, y)$  on the boundary of the room in the array plane  ($z = 2$ m). We generate $T = 600$ RIRs with random source positions. The sampling frequency is $16$ kHz. 
The input speech signal is a $1$ s clean segment randomly chosen from LibriSpeech corpus \cite{panayotov2015librispeech}. The microphone signals are obtained by convolving the RIRs with the speech signal. White noise from the Audio Set \cite{gemmeke2017audio} is added to give a $~35$ dB signal-to-noise ratio (SNR). 

\begin{figure}[t]
\centering
\includegraphics[width=7.5cm]{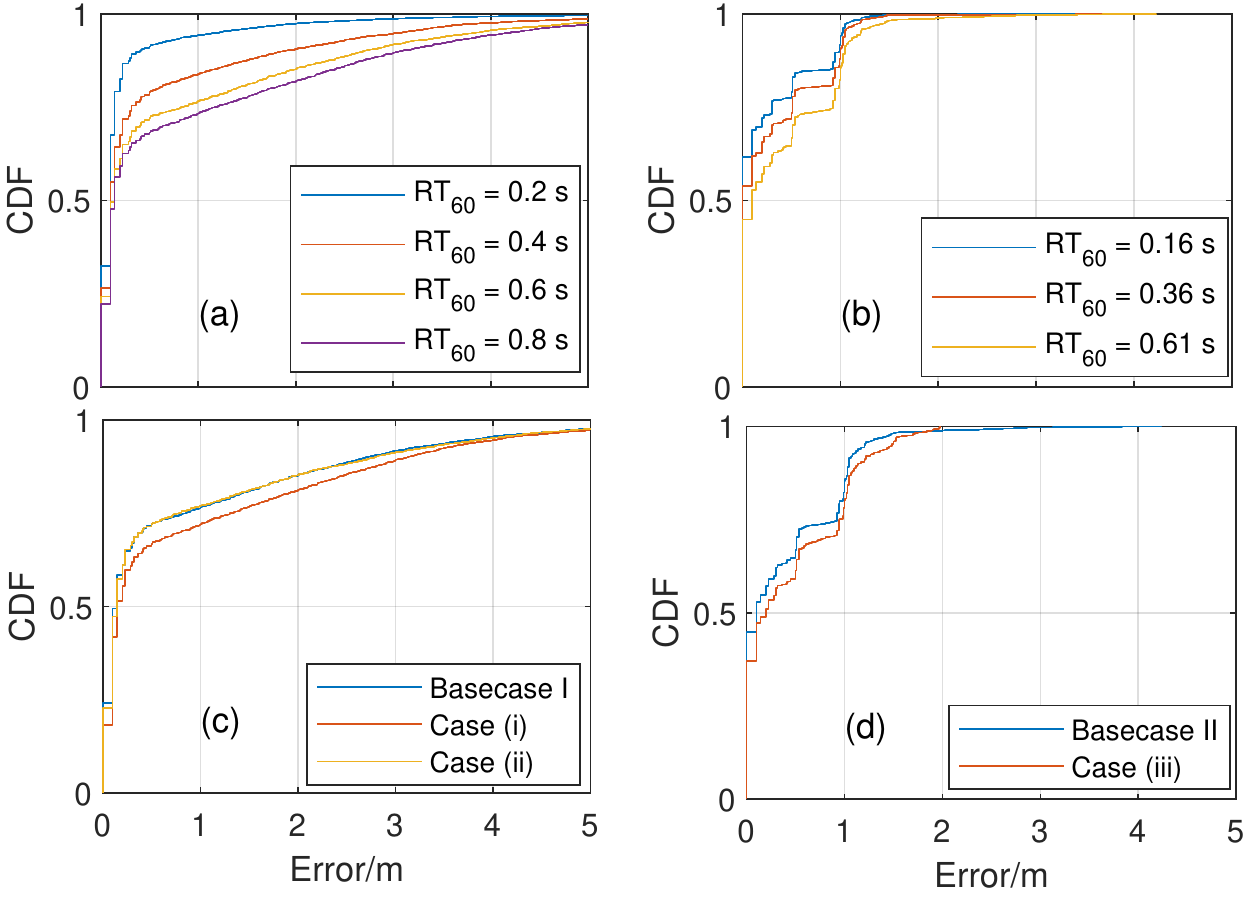}
\caption{CDF for (a) simulated data, (b) MIRD data, and generalization experiments for (c) Base Case I and (d)  Base Case II.}
\label{bbb}
\end{figure}

\subsubsection{MIRD Data}
%\textbf{MIRD Data}
These methods are also evaluated on  MIRD \cite{hadad2014multichannel} which provides  recorded RIRs for 8-microphone ULA with spacing $8$ cm for $3$  $RT_{60}$s.  We downsample  to the audio frequency from $48$ kHz to $16$ kHz.  All  reverberation times (0.16, 0.36, and 0.61 s) are applied to assess the localization performance. There are $2$ ranges ($1$ and $2$ m) and $13$ candidate DOAs, $[-90, 90^\circ]$ in $15^\circ$ steps. The sound source is  located in one of the $26$ candidate positions. We use $20$  recordings with $2$ s duration and half  female/male voices, resulting in $T = 520$ RIRs (datapoints). The noise is generated in the same way as simulation. 

%\subsection{Baseline methods}

\subsection{Parameters and implementation details}
SSLIDE is  compared with   MUSIC \cite{schmidt1986multiple}, 
SRP-PHAT \cite{dmochowski2007generalized, velasco2016proposal, yook2015fast}, SBL \cite{xenaki2018sound} and CNN \cite{chakrabarty2019multi}. The MUSIC and SRP-PHAT are implemented by Pyroomacoustics \cite{scheibler2018pyroomacoustics}. The spectrogram is used as input to train the CNN for classification. We  use $1^{\circ}$ resolution for MUSIC, SRP-PHAT and SBL in  simulations and $15^{\circ}$ for MIRD. 
%We  compare the localization DOAs using mean absolute error (MAE).

For CNN, based on the architecture suggested by \cite{chakrabarty2019multi}, we have $M - 1$ convolutional layers with kernel size $2$ ($M = 12$ for simulation and $8$ for MIRD), and $64$ filters per layer. Then, two fully connected layers (512 units for both MIRD and simulation) are added following the convolutional layers. To reduce overfitting, we apply dropout ($0.50$ dropout rate) in output layer \cite{chakrabarty2019multi}. 

For both simulated  and MIRD data, we use $N_{FFT} = 256$ with no overlap for the STFT implementation,  the number of snapshots $S_{sim} = 63$ and $S_{MIRD} = 125$. We only consider positive frequency bins, thus $F = 128$. The size for the simulation  and MIRD dataset are $C_{sim} = T \times S_{sim} = 600 \times 63 = 37, 800$ and $C_{MIRD} = T \times S_{MIRD} = 520 \times 125 = 65, 000$, separately. $\sigma$ = 0.25 (See (\ref{sigma})) is chosen for generating the target likelihoods. Fig. \ref{eg} (d) is one of the targets likelihood examples with $\sigma$ = 0.25, and we can see that it provides a sparse likelihood surface and only a small region of points that are near the target have significant values. For the simulations, the likelihood surface dimension is $101 (Y) \times 161 (X)$, and for  MIRD  $121 \times 121$. 

The model of SSLIDE is implemented by Pytorch \cite{paszke2019pytorch} with learning rate  $10^{-5}$,  batch size 32, and weight decay regularization $\lambda=5 \times 10^{-4}$, and Adam is the optimizer with weight decay $10^{-5}$. The data is split based on  $70 \%$ for training, $15 \%$ for validation and $15 \%$ for testing. The model is trained for $50$ epochs. 

\begin{table}[tb]
{
\begin{tabular}{c|c|c|c|c|c|c|c} 
\hline
                                & \multicolumn{4}{c|}{Simulated} & \multicolumn{3}{c}{MIRD }  \\ 
%\hline
$RT_{60}$/s                          & .20 & .40 & .60 & .80           & .16 & .36 & .61             \\ 
\hline
%\multirow{2}{*}{Median Error/m} & .10 & .10 & .14 & .14           & .00 & .00 & .20             \\ 
%\cline{2-8}
%                                & .14 & .28 & .71 & .72           & .00 & .00 & .30             \\ 
%\hline
{Testing}          & .24 & .56 & .77 & .90           & .23 & .29 & .39             \\ 
\hline
  Ablation                              & .54 & 1.1 & 1.8 & 1.5           & .35 & .43 & .59             \\
\hline
\end{tabular}}
%\caption{Median error and MAE performance. For median error and MAE, the first row shows the testing results, and the second row shows the results for ablation study}
\caption{MAE (m) of localization for testing data (first row) and ablation study (second row).}
\label{mmm}
\end{table}

\subsection{Results and discussions}

\textbf{Testing Performance}
The cumulative distribution function (CDF) of the localization is shown for $4$  $RT_{60}$s for the simulated data and $3$ $RT_{60}$s for MIRD data. Fig.\ \ref{bbb} (a) and (b) show the localization error distribution for simulation and MIRD  in the \textit{testing phase}. The localization error for ground-truth $(x, y)$  and predicted $(\hat{x}, \hat{y})$ location can be expressed as   $e = \sqrt{(\hat{x} - x)^{2} + (\hat{y} - y)^{2}}~$. The mean absolute error (MAE) of our approach for testing data is listed in the first row of Table \ref{mmm}.

\textbf{Comparison with Other Baseline Methods}
Since we can obtain the estimated coordinates of the sound sources, the DOA can also be accessed by simple computations. The comparison of MAE for DOA estimation with other baseline methods is in Table \ref{doaall}. From Table \ref{doaall}, we can see that our approach outperforms all  baseline methods for both the simulated and MIRD data in all levels of reverberation. For the simulated data, due to the randomness of the source positions, the localization performance of the baseline methods degrades significantly, especially when the sources are far from the sensors. In contrast, our approach can still have satisfactory performance and thus generalize across space well. For MIRD evaluation, all of the other baseline methods leverage the prior information of the candidate DOAs. In specific, for SRP-PHAT, MUSIC, and SBL, when generating the steering vectors, the distribution of DOA ($-90^{\circ}$ to $90^{\circ}$ in $15^{\circ}$ steps) is used. For CNN, it also ``knows'' that there are $13$ potential classes. Only our method does not rely on the prior information about candidate DOAs and achieves a competitive localization performance. %Among  the baseline methods, basic CNN has the best performance.   %The execution time for SBL may even exceed the whole training and testing time of the proposed DNN. 

\textbf{Ablation Study}
To validate the function of multipath alleviation decoder, we conduct ablation study which removes that decoder during the training phase. The MAEs for the testing data are listed in the second row of Table \ref{mmm}. Compared with the first row, we can see that The localization error increases when that decoder is removed for all of the cases, which verifies the role of that decoder in resolving the multipath artifacts. 

\textbf{Generalization Performance} First, we  evaluate the generalization under  perturbations of $RT_{60}$ and microphone locations (see Fig.\ \ref{bbb}(c)). The model is trained with 0.6 s $RT_{60}$ (Base Case I) but tested with the following two cases: (i)$RT_{60}$ increases to 0.7 s  (ii) the same $RT_{60}$, but microphone  spacing  increases from $2.6$  to $2.7$ cm. 

Besides, we  evaluate generalization performance  across different speech for the MIRD data (See Fig.\ \ref{bbb}(d)). The model is trained with $RT_{60} = 0.61$ s and $20$ speech signals (Base Case II) but tested with (iii) $3$ new recordings that are not used in training. We also compare the generalization performance with CNN and list the MAEs in Table \ref{generalization}. From that table, we can see that our method has a more robust performance under the perturbations of $RT_{60}$, microphone positions, and speech signal than CNN.

\begin{table}[tb]
{
\begin{tabular}{c|c|c|c|c|c|c|c} 
\hline
\multirow{2}{*}{Method} & \multicolumn{4}{c|}{$RT_{60}$/s (simulation)} & \multicolumn{3}{c}{$RT_{60}$/s (MIRD)}  \\ 
\cline{2-8}
                        & .20 & .40 & .60 & .80                        & .16 & .36 & .61  \\ 
\hline
SRP-PHAT                & 14 & 21 & 25 & 25                        & 13 & 16 & 19\\ 
\hline
MUSIC                   & 12 & 22 & 27 & 29                        & 12 & 17 & 18 \\ 
\hline
SBL                     & 7.6  & 13 & 17 & 16                        & 11 & 16 & 18 \\ 
\hline
CNN                     & 10  & 14 & 16 & 20                      & 4.6  & 8.0  & 9.8 \\ 
\hline
SSLIDE                     & 3.0  & 5.0  & 6.7 & 7.9                        & 4.3  & 5.7  & 8.1 \\
\hline
\end{tabular}}
\caption{MAE ($^\circ$) of DOA estimation for SSLIDE and other baseline methods}
\label{doaall}
\end{table}

\begin{table}
\centering
\begin{tabular}{c|c|c|c|c} 
\hline
\multirow{2}{*}{Training } & \multirow{2}{*}{Test Setup} & \multirow{2}{*}{MAE/m } & \multicolumn{2}{c}{MAE/$^{\circ}$}  \\ 
\cline{4-5}
                                &                                &                                  & SSLIDE  & CNN                 \\ 
\hline
Base Case I                     & Base Case I                    & 0.77                             & 6.66 & 15.6                \\ 
\hline
Base Case I                     & (i)                            & 0.93                             & 7.85 & 17.6                \\ 
\hline
Base Case I                     & (ii)                           & 0.78                             & 6.67 & 16.9                \\ 
\hline
Base Case II                    & Base Case II                   & 0.39                             & 8.06  & 9.81                 \\ 
\hline
Base Case II                    & (iii)                          & 0.45                             & 12.0  & 23.9               \\
\hline
\end{tabular}
\caption{Generalization performance of SSLIDE and CNN}
\label{generalization}
\end{table}

\section{Conclusions}
We developed SSLIDE, a SSL method based on a DNN with an encoder and two decoders which can localize the sources in a continous space. This  enables the DNN to simultaneously predict the locations of sound sources and mitigate multipath artifacts. Experiments indicate our method outperforms MUSIC, SRP-PHAT, SBL, and CNN in  environments with different  reverberation levels in a continuous space. The ablation study shows the importance of multipath alleviation decoder to reduce multipath and the generalization experiments show strong generalization abilities across space, perturbations of reverberation time and microphone locations, and unseen input recordings.
\label{sec:typestyle}

%\clearpage

\small
\bibliographystyle{IEEEbib}
\bibliography{strings,refs}

\end{document}